\documentclass[twocolumn,showpacs,preprintnumbers,amsmath,amssymb]{revtex4}
\usepackage{graphicx}
\usepackage{bm}
\usepackage[squaren]{SIunits}

\begin{document}
\title{Multiferroicity in Geometrically Frustrated $\alpha$-MCr$_{2}$O$_{4}$ systems (M=Ca, Sr, Ba)}

\author{Li. Zhao, Tian-Wey Lan, Kuen-Jen Wang, Chia-Hua Chien, Tsu-Lien Hung, Jiu-Yong Luo, Wei-Hsiang Chao, Chung-Chieh Chang, Yang-Yuan Chen, and Maw-Kuen Wu}\email{mkwu@phys.sinica.edu.tw}
\affiliation{Institute of Physics, Academia Sinica,  Taipei 11529, Taiwan}
\author{Christine Martin}
\affiliation{Laboratoire CRISMAT, ENSICAEN, UMR 6508 CNRS, 6 Boulevard du Mar¨¦chal Juin, 14050 Caen Cedex, France}

\date{\today}

\begin{abstract}
We have successfully synthesized three quasi-2D geometrically frustrated magnetic compounds ($\alpha$-MCr$_{2}$O$_{4}$, M=Ca, Sr, Ba) using the spark-plasma-sintering technique.  All these members of the $\alpha$-MCr$_{2}$O$_{4}$  family consist of the stacking planar triangular lattices of  Cr$^{3+}$  spins (${\rm S}=3/2$), separated by non-magnetic alkaline earth ions. Their corresponding magnetic susceptibility, specific heat, dielectric permittivity and ferroelectric polarization are systematically investigated. A long-range magnetic ordering arises below the N\'{e}el temperature (around 40K) in each member of the $\alpha$-MCr$_{2}$O$_{4}$ family, which changes to the  quasi-$120\degree$ proper-screw-type helical spin structure at low temperature. A very small but confirmed spontaneous electric polarization emerges concomitantly with this magnetic ordering. The direction of electric polarization is found within the basal triangular plane. The multiferroicity in $\alpha$-MCr$_{2}$O$_{4}$ can not be explained within the frameworks of the magnetic exchange striction or the inverse Dzyaloshinskii-Moriya interaction. The observed results are more compatible with the newly proposed Arima mechanism that is associated the $d$-$p$ hybridization between the ligand and transition metal ions, modified by the spin-orbit coupling. The evolution of multiferroic properties with the increasing inter-planar spacing (as M changes from Ca to Ba) reveals the importance of interlayer interaction in this new family of frustrated magnetic systems.
\end{abstract}

\pacs{75.85.+t, 75.10.Jm, 77.22.-d, 77.80.-e }

\maketitle

\section{Introduction}

Multiferroics, also known as ``magnetic ferroelectrics'', which exhibit the coexisting and mutual interaction of magnetism and ferroelectricity in one single phase, have been one of the most important topics that attract world-wide interest form the condensed mater physics and material science research community. The recent discovery of giant magnetoelectric coupling effect in some frustrated manganites as TbMnO$_{3}$ and TbMn$_{2}$O$_{5}$ that exhibit multiferroicity has further enhanced the interest.\cite{TbMnO3, TbMn2O5} The ferroelectricity found in these frustrated manganites is of magnetic origin, i.e., induced by complex spin orderings. These systems are thus often called ``spin-driven ferroelectrics'', as their spontaneous electric polarization arises in certain  magnetically ordered states that break the inversion symmetry. The electric polarization can be reversed or even rotated by changing the magnetic states via the application of external magnetic fields. These fascinating phenomena are of great importance  for both fundamental physics and potential technological application. \cite{Fiebig, Cheong, Ramesh, LiuJM}

The physical origin of multiferroicity is still a mater of debate. Although at present there is no commonly-accepted unified microscopic theory, several possible mechanisms have been proposed to provide the understanding of some certain multiferroic systems.\cite{LiuJM} One simple mechanism for multiferroicity is the exchange striction in certain (quasi-)collinear commensurate spin configurations with inequivalent magnetic ions.\cite{Cheong, LiuJM} For example, the  inequivalent nearest-neighbor exchange striction in a 1D  collinear  $\uparrow\uparrow\downarrow\downarrow$ spin chain system can break the space inversion via shifting ions away from centrosymmetric positions, then induce a spontaneous electric polarization along the spin chain, as observed in Ca$_{3}$(Co, Mn)O$_{6}$.\cite{CCMO}
But the spin configurations in most of the known multiferroic materials are much more complex at low temperatures, which are usually noncollinear and incommensurate. Thus, alternative mechanisms were proposed for the non-collinear spin ordering.

According to the inverse Dzyaloshinskii-Moriya (DM) mechanism \cite{DM} or the equivalent spin-current model (proposed by Katsura, Nagaosa and Balatsky, also known as KNB model),\cite{KNB}  the two neighboring atomic sites with mutually canted spins will induced a local electric polarization via spin-orbit interaction. This induced local polarization can be formulated as  ${\bm P}_{ij}=A \hat{\bf {e}}_{ij} \times ( {\bm S}_i  \times {\bm S}_j )$, where the coupling coefficient $A$ is determined by the spin-orbit coupling and exchange interactions.  $\hat{\bf {e}}_{ij} $ is the unit vector connecting sites $i$ and $j$. This expression is also consistent with the phenomenological theory based on symmetry consideration,\cite{Mostovoy}  and has explained qualitatively the origin of the ferroelectricity in several helical magnetic systems as TbMnO$_{3}$,\cite{DM}  LiCu$_{2}$O$_{2}$, \cite{LiCu2O2}  etc.

Nevertheless, it is noted that the above mechanisms cannot predict any electric polarization for a class of so-called ``proper-screw-type'' helimagnets, in which the magnetic modulation vector is parallel to the spin helicity (defined as ${\bm S}_i  \times {\bm S}_j $). However, multiferroicity have been discovered in  several delafossite-like magnetic materials with hexagonal lattices, like CuFeO$_{2}$,\cite{CuFeO2}  ACrO$_{2}$ (A=Cu, Ag),\cite{ACrO2} etc. These materials are all  low-dimensional triangular-lattice antiferromagnets with strong geometrical magnetic frustration. The corresponding spontaneous electric polarization arises concurrently with the emerging proper-screw magnetic ordering at low temperature. To account for these experimental observations, Arima \cite{Arima} and Jia et al \cite{Jia} proposed a new theoretical scheme recently, in which the certain type of proper-screw spin orderings are capable of inducing macroscopic electric polarization through the variation in the metal-ligand hybridization with spin-orbit coupling. However, till now, a unified understanding and quantitative prediction of multiferroicity still lacks in the present research.

On the other hand, the research of the geometrically magnetic frustration in the strongly-correlated electronic systems has been of long interest. \cite{Ramirez}   Usually, the short range orderings, strong fluctuations, rich phase diagrams and novel critical phenomena  arise at low temperatures in these systems, providing a fertile ground for physicists. The competing neighboring magnetic interactions lead to the extra degeneracy in ground state over and above the non-frustrated systems, which give rise to the emerging of new physics. The anisotropy of magnetic ions (of easy-axis or easy-plane type), the interlayer coupling, etc, also play important roles in real systems. The simplest form of the magnetically geometrical frustration can arise from a 2D triangular arrangement of magnetic ions that are coupled antiferromagnetically with each other. Theoretically, for an ideal 2D triangular lattice of identical classical vector spins, there are degenerate solutions for the lowest energy of the system in which each given spin vector arranged at $120\degree$ to its nearest neighbors. The highly degenerate ground states lead to exotic physical magnetic behaviors, such as short-range ordered spin liquid states, or long-range chiral magnetic orderings.\cite{Collins}  In real antiferromagnets with quasi-2D triangular lattice, the intralayer structural distortions from an ideal regular triangular lattice, the interlayer magnetic exchange interactions, and the anisotropy of magnetic ions can act as important additional terms, leading to many different types of magnetic structures. For example, the interlayer interactions can stabilize the long-range magnetic ordering at higher temperatures, and the strong single-ion anisotropy favors the collinear ordering in the ground state. There are still many open questions about the physical behavior of geometrically frustrated magnets.

In this paper, we investigate the $\alpha$-phase MCr$_{2}$O$_{4}$ family systematically, where M stands for different alkaline earth metal (M=Ca, Sr, or Ba).  As shown in Fig. 1, the $\alpha$-phase MCr$_{2}$O$_{4}$ has a quasi-2D layered structure with triangular CrO$_{2}$ sheets, which are made of edge-sharing CrO$_{6}$  octahedra. These stacking CrO$_{2}$  sheets are well separated by nonmagnetic ions of alkaline earth metal. The different M atoms primarily manifest in different separation ($d$) between the triangular CrO$_{2}$  sheets. With increasing cationic radius, $d$ increase markedly, i.e., $d$=5.53 \angstrom, 5.82 \angstrom and 6.14 \angstrom~ for M=Ca, Sr, and Ba, respectively.

\begin{figure}
\includegraphics[width=0.4\textwidth]{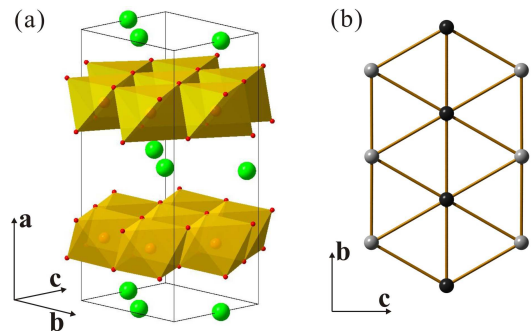}
\caption{\label{f1}
(Color online) (a) Crystal structure of $\alpha$-MCr$_{2}$O$_{4}$. The large green spheres stand for the alkaline earth ions (M=Ca, Sr, Ba). And the small red balls represent the oxygen ions, forming the octahedra in which are chromium cations. (b) Projection along the a-axis of a single triangular  CrO$_{2}$ layer with only the Cr$^{3+}$  network. Two kinds of the inequivalent Cr$^{3+}$  sites are plotted as the dark and light gray spheres respectively.}
\end{figure}

In each CrO$_{2}$  layer, the neighboring Cr$^{3+}$-Cr$^{3+}$  distance is short (close to 3 \angstrom), resulting a strong direct antiferromagnetic exchange interaction, while the intra-layer superexchange is known to be quite weak since the Cr$^{3+}$-O-Cr$^{3+}$  bonding angle is around 95$\degree$.\cite{JPSJ1970} The exchange interaction between interlayer Cr$^{3+}$  ions is quite weak due to much larger distance between layers. The structure of $\alpha$-MCr$_{2}$O$_{4}$ resembles the hexagonal delafossites closely, with a slight rectangular distortion in the CrO$_{2}$  layers from the ideal regular triangular lattice. This kind of distortion results an over-all orthorhombic symmetry in $\alpha$-MCr$_{2}$O$_{4}$ with two inequivalent Cr$^{3+}$  sites as shown in Fig1. (b).

Although $\alpha$-phase MCr$_{2}$O$_{4}$ had been synthesized and structurally characterized dozens of years ago, \cite{1974a,1974b,1974c}  their physical properties at low temperature have not been investigated until recently. Chapon et al \cite{Chapon} first investigated the magnetism and magnetic structure of polycrystalline $\alpha$-CaCr$_{2}$O$_{4}$. A long-range ordered helical magnetic structure in $\alpha$-CaCr$_{2}$O$_{4}$ occurs below the  N\'{e}el temperature ($T_N$) of about 43K with an incommensurate propagation vector $\bf{k}$=(0, $\sim$1/3, 0). The neutron diffraction experiment revealed that the spins in CrO$_{2}$  layer lie in the $ac$ plane, perpendicular to the  $\bf{k}$ vector. This is a typical ``proper-screw-type'' helimagnetic structure. And the angle between the neighboring spins is around $120\degree$, which is very close to the theoretical expected value for an ideal 2D triangular antiferromagnetic lattice, in spite of the orthorhombic distortion in $\alpha$-CaCr$_{2}$O$_{4}$. Later, a neutron measurement on crystal samples further confirmed this quasi-$120\degree$ helical spin picture.\cite{Toth}

In 2011, Cava's group \cite{Dutton}  synthesized the polycrystalline $\alpha$-SrCr$_{2}$O$_{4}$, a Sr version of $\alpha$-MCr$_{2}$O$_{4}$ with an expanded spacing between the CrO$_{2}$  layers and also a slightly larger intra-layer orthogonal distortion. The powder neutron diffraction measurements shows a very similar incommensurate helical spin ordering arising below almost the same magnetic transiton temperature as in $\alpha$-CaCr$_{2}$O$_{4}$. Replacing Ca$^{2+}$ ions with larger Sr$^{2+}$ ones seemed incapable of bringing about significant changes in the physical properties in this class of materials. Further measurements on the temperature-dependent structure parameters revealed a  anomalous slight inflection in the lattice parameter perpendicular to CrO$_{2}$  layers around $T_N$, suggesting a possible magneto-structural coupling. The quadratic magnetoelectric effect was also predicted based on a Landau symmetry analysis. Recently, the neutron diffraction measurements on bulk samples of $\alpha$-BaCr$_{2}$O$_{4}$  by Hardy et al disclosed a similar long-range incommensurate helimagnetic phase below $T_N$, very close to the above quasi-$120\degree$ spin structure, which will be discussed else where.\cite{Hardy}

In addition, Singh et al \cite{Singh} reported the multiferroicity existing in the bulk samples of $\alpha$-CaCr$_{2}$O$_{4}$. The weak ferroelectricity occurs concomitantly with the incommensurate helical magnetic ordering below $T_N$ of $\alpha$-CaCr$_{2}$O$_{4}$. But no further microscopic mechanism is given. Based on the structural similarity, we can reasonably expect that the similar multiferroicity also exists in the other two analogues in the $\alpha$-MCr$_{2}$O$_{4}$ family. Therefore, the systematical study on this family is clearly needed to clarify their microscopic physical origin of the multiferroicity and search for new promising multiferroic materials among the geometrically frustrated triangular systems.

Herein,  we report the study of all the three member compounds (M=Ca, Sr and Ba) using samples of high quality. Based on the measurements of the magnetic susceptibility, specific heat and especially, dielectric and ferroelectric properties, our work confirms the spin-driven ferroelectricity in this family. In the following, we first focus on $\alpha$-SrCr$_{2}$O$_{4}$ (the middle member in this family) and then systematically compare it with the other two members ($\alpha$-phase CaCr$_{2}$O$_{4}$ and BaCr$_{2}$O$_{4}$). We also provide our view on the possible physical mechanisms for the observed multiferroicity.

\section{Experimental details}

Our polycrystalline samples of $\alpha$-MCr$_{2}$O$_{4}$ (M=Ca, Sr, Ba) were prepared by the spark-plasma-sintering(SPS) method. All these compounds own very high melting-points($T_{mp}$). For M= Ca and Sr, $T_{mp}>2000\degreecelsius$. As for Ba, $T_{mp}>1600\degreecelsius$.\cite{1974a,1974b,1974c}   A rather high sintering temperature ($>1400\degreecelsius$) is necessary to prepare pure $\alpha$-phase samples, because of a two-phase competition in the MCr$_{2}$O$_{4}$ system. The $\alpha$-phase MCr$_{2}$O$_{4}$ favors high sintering temperature while another $\beta$-phase is prone to appear in low temperature.\cite{beta} A controlled atmosphere is also necessary to prevent low-valence  Cr$^{3+}$  ions from being further oxidized. Compared with the conventional solid-state sintering synthesis, the new powerful SPS technique can  be used to prepare many refractory materials productively with high  density and it also facilitates a very high heating and cooling rate to avoid  $\beta$-phase as impurity in our samples (a rate of about 200\degreecelsius/min is adopted in our experiments).

The starting materials of MCr$_{2}$O$_{4}$ for SPS processing is prepared by a conventional ceramic synthesis route. The stoichiometric mixture of  Cr$_{2}$O$_{3}$ and MCO$_{3}$ with high purity was sintered at 1050$\degreecelsius$ under a dynamic high vacuum (below 10$^{-6}$ Torr) for 72 hours with several intermediate grindings. Then the precursor was loaded in to a graphite die, consolidated by SPS processing. Sample is kept at 1400-1500$\degreecelsius$ for at least 20min under an uniaxial pressure of 40 MPa and a very large dc current (over 1000 A) is applied, using a Dr. Sinter 1500 SPS machine (SPS SYNTEX Inc.). The final products are very dense pellets with 15 mm in diameter and around 5 mm in thickness. The high quality of our samples is tested  by x-ray diffraction, and later confirmed by our magnetization and specific heat measurements. No $\beta$-phase and other impurities as Cr$_{2}$O$_{3}$ were found. The details about the sample preparation by SPS will be reported elsewhere.

The magnetic properties of our samples were measured using a SQUID VSM magnetometer (Quantum Design inc.). The measurements of specific heat were carried out using a standard thermal relaxation calorimetric method in a commercial Physical Property Measurement System (PPMS, Quantum Design Inc.).

For dielectric measurements, the highly dense pellets of $\alpha$-MCr$_{2}$O$_{4}$ prepared by SPS were cut into a rectangle plate first and then polished to a thin sheet with thickness of 0.1- 0.3 mm. We use silver paint attached to both sides as electrodes to form a parallel-plate-capacitor. The samples are glued on the cryogenic stage of our homemade probe, and connected to a high-precision capacitance meter via low-loss coaxial cables. The main sources of error such as residual impedance in the whole circuit have been carefully compensated.

The  electric polarization ($P$) is obtained from the integration of the corresponding pyroelectric current. We first polarize the specimens with a static electric field of 600-1200 kV/m during the cooling process, then remove the electric field and short-circuit the sample at low temperature for about one hour to remove the possible stray charge carriers. The pyroelectric current was measured during the warming process at different heating rates (1-4 K/min).

Besides the bulk samples prepared by SPS, we also try to grow the single crystals using a high temperature optical floating zone furnace (FZ-T-12000-X-VPO, Crystal systems Corp. Japan), which is equipped with four elliptical mirrors and four 3 kW Xenon lamps.
At the moment, we can only successfully grow large crystals of $\alpha$-SrCr$_{2}$O$_{4}$. The shiny flake-like crystals with large area (usually exceeding 4 mm$^2$) can be easily cleaved from the boule, and they are highly $a$-axis oriented. Only the dielectric properties along $a$-axis can be measured since these flake-like crystals are very thin (less than 0.5mm). X-ray phi-scan measurements on a four-circle diffractometer reveal a highly 60\degree-twinning structures in these crystals, since the in-$bc$-plane orthogonal structure in $\alpha$-SrCr$_{2}$O$_{4}$ is very close to a ideal hexagonal lattice. Therefore the in-plane anisotropic physical properties (along $b$- or $c$-axis) cannot be distinguished at present. The crystal growth, together with the corresponding characterization of its structural and physical properties will be reported in detail elsewhere.

\section{Experimental Results}

\subsection{$\alpha$-SrCr$_{2}$O$_{4}$}
Firstly, we present the results on $\alpha$-SrCr$_{2}$O$_{4}$, including its magnetic susceptibility, specific heat and especially, dielectric and ferroelectric properties of the bulk samples. Some results on crystal samples are also presented.

\subsubsection{Magnetic properties}
We measured the magnetic susceptibility ($\chi$) of bulk $\alpha$-SrCr$_{2}$O$_{4}$ samples in several magnetic fields. The $\chi(T)$ curves measured in both the zero-field-cooling(ZFC)  and the field-cooling(FC) processes almost merge together. The linear behavior in the field-dependent magnetization (M-H curve, not show here) is observed up to H=7T (the limit of our SQUID VSM magnetometer) at several different temperatures below and above the magnetic transition temperature, indicating that the basic magnetic structure in $\alpha$-SrCr$_{2}$O$_{4}$ holds robust even in a strong field and no possible metamagnetic transition such as spin flop or flip occurs.

\begin{figure}
\includegraphics[width=0.4\textwidth]{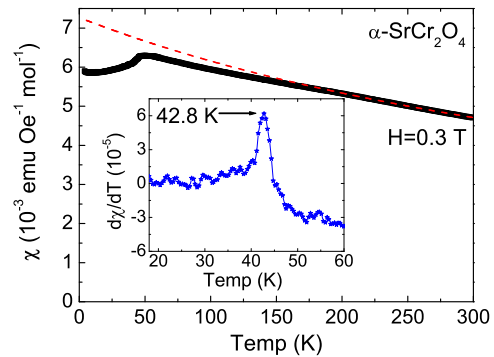}
\caption{\label{f2}
(Color online) The temperature dependence of magnetic susceptibility ($\chi$). The red dash line is the Curie-Weiss fit. And the corresponding temperature derivative (${\rm d}\chi/{\rm d}T$) is shown in the inset.
}
\end{figure}

The mainframe in Fig. 2 shows the over-all temperature dependence of magnetic susceptibility measured on a typical bulk sample in H=0.3T. Upon cooling from room temperature, $\chi(T)$ of $\alpha$-SrCr$_{2}$O$_{4}$  increases gradually. At low temperature, $\chi(T)$ shows a broad hump with a maximum around 50K, reflecting a typical characteristic of a fluctuating short-range antiferromagnetic ordering in low dimensional systems. As temperature decreases further, $\chi(T)$ drops abruptly at about 43K, indicating the emerging of a long-range antiferromagnetic ordering. For clarification, the corresponding temperature derivative (${\rm d}\chi/{\rm d}T$) is plotted in the inset. The N\'{e}el temperature for this antiferromagnetic transition, $T_N$, can be accurately determined from the peak in ${\rm d}\chi(T)/{\rm d}T$. $T_N$ is about 42.8K, which is consistent with reported results.\cite{Dutton}

Above $T_N$, the behavior of $\chi(T)$ does not fit well with the Curie-Weiss formula ($\chi(T)=C/(T-\theta_{CW})$), except in the high temperature range between 200K and 300K. The fitted curve (dashed line) is also plotted in Fig. 2. The acquired Curie-Weiss temperature, $\theta_{CW}$, is -552(.5)K, much larger than $T_N$. The $f$ value (the quantity defined as $\theta_{CW}/T_N$, an empirical measure of frustration) is relatively large (close to 13), indicating a strong magnetical frustration in $\alpha$-SrCr$_{2}$O$_{4}$. The effective magnetic moment (calculated from the fitted C parameter) is $\mu_{eff}$=4.0(2) $\mu_B$, consistent with the previous results.\cite{Dutton}  However, the expected moment value for the paramagnetic ${\rm S}=3/2$ Cr$^{3+}$  ion is 3.87$\mu_B$. The difference may be attributed to the narrow temperature range over which the Curie-Weiss fitting was carried out. The fit to the high temperature $\chi(T)$ (T changes from 300 to 1000K) in $\alpha$-CaCr$_{2}$O$_{4}$ has given a quite close value to the expected moment.\cite{Toth} Therefore, for a better Curie-Weiss fitting, the measurement of  $\chi(T)$ well above room temperature is needed in our further work.

\subsubsection{Heat capacity measurements}

The further evidence of the long range antiferromagnetic ordering in $\alpha$-SrCr$_{2}$O$_{4}$ comes from the measurement of specific heat ($C_p$). The $C_p$(T) shows a sharp $\lambda$-shaped anomaly at $T_N$ (see Fig. 3). The over-all behavior agrees well with the published data on the $\alpha$-SrCr$_{2}$O$_{4}$ bulk sample, which was prepared via a conventional method by Dutton et al.\cite{Dutton} It is noticeable that the weak anomalous feature at 35K reported in their work (Fig. 3 in Ref. \onlinecite{Dutton}) is not observed in our data, indicating that the 35K anomaly may come from some unknown impurities in their sample or possible experimental artifact.

\begin{figure}
\includegraphics[width=0.4\textwidth]{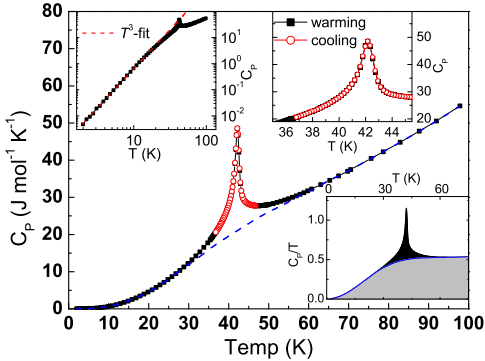}
\caption{\label{f3}
(Color online) Temperature dependence of the specific heat, $C_p(T)$, for $\alpha$-SrCr$_{2}$O$_{4}$. The main frame shows the data measured in warming (solid square) and cooling processes (open circles). The dashed blue line represents the background to be subtracted to estimate the change in entropy by the anomaly peak around $T_N$. The corresponding $C_p(T)/T$ curve is shown in the bottom right inset. The upper right inset focuses $C_p(T)$ in the region around the magnetic transition. The overall $C_p(T)$ is also plotted on the log-log scale in the upper left inset, with a Debye-type $T^3$ power fit (red dash line) to the low temperature part of $C_p(T)$.
}
\end{figure}

The total specific heat consist of two main contributions, one from lattice vibrations (phonons) and the other from magnetic excitations (magnons), i.e $C_p\approx$$C_{phonon}+C_{magnon}$.   For magnetic systems of transition metal with nontrivial exchange, these two contributions are usually comparable, especially in low low temperature, as summarized by Gopal.\cite{Cp} As shown in upper-left inset in Fig. 3. A simple $T^{3}$ power fit (thin solid line) agrees well with the experimental data below 20K.  It is well-known that $C_{phonon}$ obeys the Debye $T^{3}$ power law at low temperature. Therefore, the magnetic contribution should also obey the similar $T^{3}$ law, i.e. $C_{magnon} \propto T^{3}$. The  $T^{3}$-power-like behavior for $C_{magnon}$ confirms the long-range antiferromagnetic ordering in $\alpha$-SrCr$_{2}$O$_{4}$, \cite{Cp}  and excludes other possibilities of ferro- and ferri- magnetism that both obey the $T^{3/2}$ law.\cite{Cp}

At present, it is hard to extract solely the magnetic part in $C_p$(T) accurately due to the absence of non-magnetic isostructural analogues to estimate the contribution of the lattice vibration to the total specific heat. To just estimate roughly the entropy removed by the long-range antiferromagnetic ordering which is signaled by the peak of $C_p(T)$ around $T_N$, we use the power series expansion fit to the high temperature($>$60K) and low temperature($<$25K) regimes as the background(the blue dash line shown in Fig. 3). The entropy, obtained from integrating the anomaly part in $C_p(T)/T$ (the dark grey shadow in the bottom inset), is 2.9 \joulepermolekelvin, only 12.6\% of the full magnetic entropy associated with $2Rln(2S+1)\approx23.05 \joulepermolekelvin$ as expected for a ${\rm S}=3/2$ spin system. Most of remaining  magnetic entropy is considered to be removed by the short-range ordering well above $T_N$, as usually observed in the low-dimensional frustrated systems.\cite{Ramirez}

There is still no consensus yet regarding the order of the the magnetic transition in $\alpha$-MCr$_{2}$O$_{4}$. cite{Chapon,Toth,Dutton,Singh}  We further checked whether there exists any hysteresis in specific heat during the heating and cooling cycle around around $T_N$.  For bulk $\alpha$-SrCr$_{2}$O$_{4}$, the two set of data merges well within the limit of our experimental resolution (the details are shown in the upper right inset in Fig. 3). The same results are also observed on other samples with M=Ca, Ba (not shown here).  Considering the fact that people didn't observe any corresponding structural transition at $T_N$ till now, our results support the nature of second order in this antiferromagnetic transition at $T_N$.

\subsubsection{Bulk dielectric and pyroelectric measurements}

Fig. 4(a) shows the raw capacitance data, which is proportional to dielectric constant ($\epsilon$), measured on a typical sample of bulk $\alpha$-SrCr$_{2}$O$_{4}$ in zero field during the warming process. The different testing frequencies (ranging from 30 kHz to 1 MHz) are used to exclude the possible non-intrinsic artifacts. Similar dielectric behaviors have been observed for all the testing frequencies in our measurements.

\begin{figure}
\includegraphics[width=0.4\textwidth]{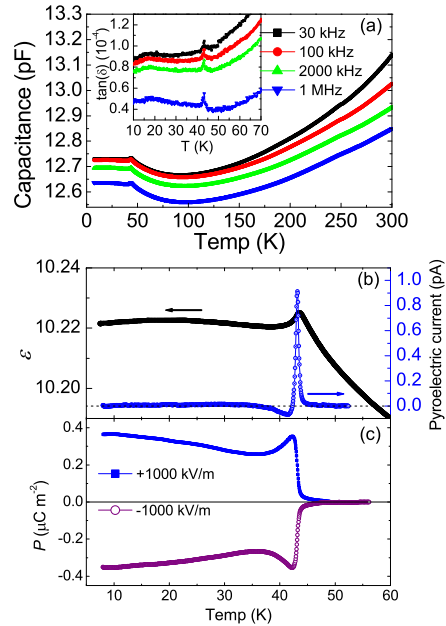}
\caption{\label{f4}
(Color online) (a) Raw capacitance data of a typical $\alpha$-SrCr$_{2}$O$_{4}$ sample measured at different frequencies in zero field. The corresponding dielectric loss (tan$\delta$) is shown in the inset. (b) Temperature dependence of the dielectric constant (black) measured at a frequency of 100kHz and pyroelectric current (blue) measured in the warming process after poling with a electric field of 1000 kV/m. The dashed line represents zero pyroelectric current. (c) Electric polarization ($P$) vs. temperature, using both positive (blue) and negative(cyan) poling electric fields.
}
\end{figure}

At high temperature, the dielectric constant for $\alpha$-SrCr$_{2}$O$_{4}$ decreases gradually upon cooling, which is typical for most insulating materials. But below roughly at 100K, the dielectric constant turns around to become slightly increasing with decreasing temperature. The anomalous increase in $\epsilon(T)$ is likely related to the short-range antiferromagnetic ordering emerging well above $T_N$ via magnetoelectric correlations. It has been observed in other systems as TeCuO$_{3}$ that the short-range magnetic correlations
 affect on dielectric properties.\cite{Lawes}

The most remarkable feature is the small lambda-like dielectric anomaly peaked at the $T_N$, concurrently with the emerging of a long range magnetic ordering, which has been observed in the magnetization and heat capacity measurements. Correspondingly, a very weak peak arises in the correspondingly dielectric loss (tan$\delta$) at the same temperature (shown in the inset), indicating the signature of a weak but corroborative intrinsic ferroelectric transition at $T_N$.

Besides adopting different testing frequencies, we have also confirmed the anomalous features in $\epsilon$(T) by repeating the temperature cycles and on samples of different batches (the same testing procedures were carried out on other samples like $\alpha$-phase  CaCr$_{2}$O$_{4}$ and BaCr$_{2}$O$_{4}$ which are discussed later). Therefore, possible extrinsic factors such as trapped interfacial charge carriers and thermal degradation of bulk sample can be excluded. The observed anomalies indicate an intrinsic electric transition at $T_N$.  For convenience, we only use the data measured at 100 kHz in our following discussions.

The weak ferroelectricity in $\alpha$-SrCr$_{2}$O$_{4}$ is further confirmed in our measurements of electric polarization using the pyroelectric method. Fig. 4(b) shows the pyroelectric data, along with the corresponding dielectric constant,  measured on a typical bulk sample. A quite small (below 1 pA) but very sharp peak emerges as the temperature approaches $T_N$. The current drops to zero as $ T \ge T_N$. The $P$ is acquired by integrating the pyroelectric current from above $T_N$. A clear development of a spontaneous electric polarization $P$ below the magnetic ordering temperature is shown in Fig. 4(c). Below $T_N$,  $P$ can be inverted with the opposite poling electric field, proving its ferroelectric nature. At T = 10 K, $P$ saturates at about  $0.36 \mu C/m^{2}$, a very small value which is several orders smaller than most known multiferroic materials,\cite{LiuJM} but of the same order as the published data on isostructural $\alpha$-CaCr$_{2}$O$_{4}$ by Singh et al.\cite{Singh}

Another important feature is from the temperature-dependent pyroelectric current. As temperature increases, a small negative pyroelectric hump arising around 40K, as a precursor close to the sharp peak just below $T_N$, as shown in Fig. 4(b).  Correspondingly, as temperature decreases from above $T_N$, $P$ first goes up rapidly from zero to a maximum just below $T_N$ , then  $P$ decreases as the result of the emerging negative pyroelectric current.  $P$ continues to decreases to a minimum at about 37K, at which the negative pyroelectric current vanishes. Upon further cooling,  $P$ increase very slowly and saturates at lowest temperature. The decreasing in  $P$ on cooling in a narrow temperature range (between 37K and 42K) is rarely observed in conventional ferroelectric materials. It is noted that such an anomaly has been observed on the $\alpha$-CaCr$_{2}$O$_{4}$ samples by Singh et al without giving the corresponding explanation.\cite{Singh}  The possible artifacts such as interfacial trapped carriers have been carefully excluded by reversing poling electric field and adopting different heating rate in our measurements. Jodlauk et al  have also observed the similar negative pyroelectric currents on some multiferroic pyroxenes.\cite{Jodlauk}  Although its origin is not yet clear, the observed anomaly is unlikely attributed to ferroelectricity. One possible explanation is the existence of the strong antiferroelectric correlation between the CrO$_{2}$  layers, which will be discussed later.

\subsubsection{Magnetoelelctric coupling in bulk $\alpha$-SrCr$_{2}$O$_{4}$}

We further investigated the magentoelectric coupling effect in $\alpha$-SrCr$_{2}$O$_{4}$ in magnetic field up to 9T, which is parallel with the plate-like sample (i.e. H$\bot$E configuration). We also tested the H$\parallel$E measuring configuration in our experiments. No noticeable difference was observed, since the intrinsic anisotropy is possibly cancelled out in our polycrystalline samples.

\begin{figure}
\includegraphics[width=0.4\textwidth]{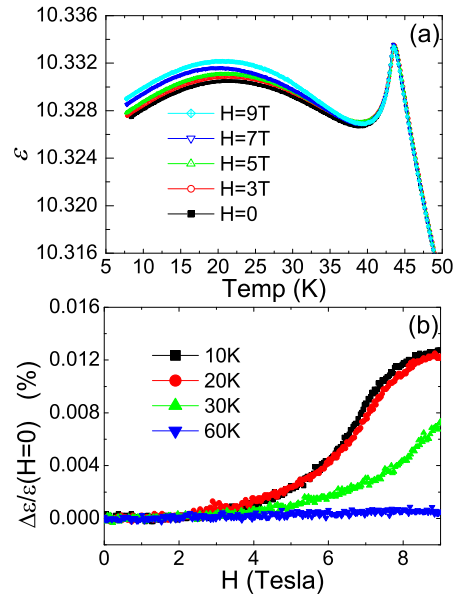}
\caption{\label{f5}
(Color online) (a) Temperature dependence of dielectric constant ($\epsilon$) for $\alpha$-SrCr$_{2}$O$_{4}$  in different applied magnetic fields (H=0-9T). (b) Magnetic field dependence of $\epsilon$ at several temperatures above (T=60K) and below (T=10, 20, 30K) the magnetic transition temperature.
}
\end{figure}

The $\epsilon$(T) under different field (H=0-9T) is measured (see Fig.5 (a)). No noticeable temperature shift of the antiferromagnetic transition (deduced from the peak of dielectric anomaly) is observed, confirming the robustness of the helimagnetic structure in $\alpha$-SrCr$_{2}$O$_{4}$. At high temperatures, $\epsilon$(T) is almost not affected by H. Below $T_N$, $\epsilon$(T) increases quite slightly upon increasing H. The effect on $\epsilon$(T) by H is vanishing small (an order of $10^{-5}$) and it develops slightly with further cooling. We also measured $P(T)$ under different field (not shown here), the difference is negligible within the experimental resolution of our setup.

To quantify this magnetoelectric coupling effect more precisely, we measured the magnetoelectric coefficients (MC, defined as ($\epsilon$(H)- $\epsilon$(H=0))/$\epsilon$(H=0)) at several temperature points above and below $T_N$ (see Fig. 5(b)). Above $T_N$, MC is almost zero, suggesting the magnetoelectric coupling in $\alpha$-SrCr$_{2}$O$_{4}$ is associated only with the long-range magnetic ordering below $T_N$. At $T < T_N$, MC is positive and increases with decreasing temperature. And the corresponding magnitude is very small, only reaching $10^{-4}$ at the highest field H=9T, much smaller than most of the known multiferroic materials.\cite{LiuJM}

At each temperature point below $T_N$, MC increases with increasing H, approximately proportional to the square of the field, except the saturating part in the high field region(H $\ge$ 8T). This quasi-quadratic magnetoelectric effect is consistent with the empirical analysis based on the analysis of the Landau free energy and symmetry consideration by Chapon and Singh et al.\cite{Chapon, Singh}

\subsubsection{Results on crystal samples of $\alpha$-SrCr$_{2}$O$_{4}$}

As aforementioned, our  flake-like crystal samples of $\alpha$-SrCr$_{2}$O$_{4}$ are highly $a$-axis oriented, with quasi-60\degree twinning in $bc$ plane. Therefore, only the dielectric constant and electric polarization along a-axis ($\epsilon_a$ and $P_a$) can be measured  at present. In low temperature, the corresponding magnetic susceptibilities in H$\bot a$ and H$\parallel a$ have been measured (not shown). And both behave very similar as the bulk sample, which is consistent with the published data on the isostructural  crystals of $\alpha$-CaCr$_{2}$O$_{4}$ by Toth et al.\cite{Toth}

\begin{figure}
\includegraphics[width=0.4\textwidth]{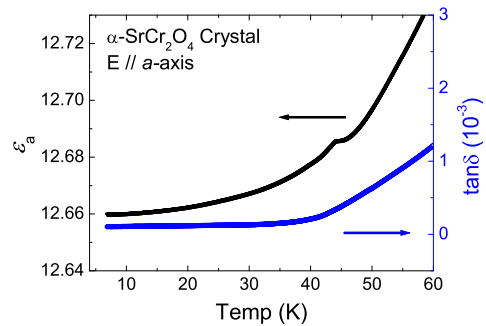}
\caption{\label{f6}
(Color online) (a) Temperature dependence of $a$-axis dielectric constant($\epsilon_a$) and corresponding loss (tan$\delta$) measured on a flake-like $\alpha$-SrCr$_{2}$O$_{4}$ crystal sample grown by optical floating zone method. The testing electric field E is along $a$-axis.
}
\end{figure}

As shown in Fig. 6, in zero field, $\epsilon_a$(T) decreases gradually on cooling. Around $T_N$, only a very weak hump is observed. This weak anomaly in $\epsilon_a(T)$ is associated with the multiferroic transition, which has been observed on bulk samples. No anomaly at $T_N$, can be visible in the corresponding dielectric loss. We also measured $\epsilon_a(T)$  under different H, the influence in $\epsilon_a(T)$ by H is negligible. In our pyroelectric measurements, no net electric polarization ($P_a$) is observed within our experimental resolution, which strongly contrasts against the results on the bulk samples. This result suggests the absence of ferroelectricity along $a$-axis in $\alpha$-SrCr$_{2}$O$_{4}$. Thus, the intrinsic ferroelectric polarization, arising from the magnetic ordering in $\alpha$-SrCr$_{2}$O$_{4}$, must lie in $bc$-plane.

At present, our crystal samples and the experimental setup does not allow us to determine the exact direction of in-$bc$-plane polarization. According to the analysis of magnetic symmetry, \cite{Arima}  the proper-screw-type spin ordering, which emerges below $T_N$, breaks the spatial inversion and mirror operation, and makes the only allowed 2' operation axis along the the magnetic modulation vector ($b$-axis in $\alpha$-SrCr$_{2}$O$_{4}$). So the polar order should appear in this direction. The growth of large untwinned single crystals is necessary in our further studies on its anisotropic properties.

\subsection{$\alpha$-phase  CaCr$_{2}$O$_{4}$ and BaCr$_{2}$O$_{4}$}

The other two members in the family of $\alpha$-MCr$_{2}$O$_{4}$ with different alkaline earth elements (M=Ca, Ba) have also been synthesized by the SPS method. On the whole, their physical properties are quite similar to those of $\alpha$-SrCr$_{2}$O$_{4}$.

Fig. 7 (a) and (c) show $\chi(T)$ and the corresponding temperature derivatives (${\rm d}\chi/{\rm d}T$) for bulk samples of $\alpha$-phase  CaCr$_{2}$O$_{4}$ and BaCr$_{2}$O$_{4}$, respectively. The antiferromagnetic humps in $\chi(T)$ are clearly observed on both compounds, very similar to that of $\alpha$-SrCr$_{2}$O$_{4}$. The neutron diffraction measurements on the powder and single crystal samples of $\alpha$-CaCr$_{2}$O$_{4}$ \cite{Chapon, Toth} have revealed a long-range incommensurate proper-screw-type helimagnetic ground state, which is almost identical to that observed in $\alpha$-SrCr$_{2}$O$_{4}$.\cite{Dutton}  An analogous helical spin picture  in $\alpha$-BaCr$_{2}$O$_{4}$ was just proposed based on the recent neutron powder diffraction by Hardy et al.\cite{Hardy} The high similarity of magnetic structures in the $\alpha$-MCr$_{2}$O$_{4}$ family accounts for their analogous magnetic properties, which comes from the basal structure of the quasi-2D triangular CrO$_{2}$ layers.

The clear kink in $\chi(T)$ labels the emerging of a long-range magnetic ordering. The corresponding $T_N$ is determined accurately according to the sharp peak in ${\rm d}\chi/{\rm d}T$. $T_N$ of $\alpha$-CaCr$_{2}$O$_{4}$ is 43.0K, very close to that of $\alpha$-SrCr$_{2}$O$_{4}$ (42.8K). However, for $\alpha$-BaCr$_{2}$O$_{4}$, $T_N$ drops to 39.5K, a few Kelvins lower than the other two members. The difference can be easily understood, considering the much larger interlayer spacing in $\alpha$-BaCr$_{2}$O$_{4}$ (6.14 \angstrom) than the other two members (5.53\angstrom~for $\alpha$-SrCr$_{2}$O$_{4}$ and  5.82\angstrom~for $\alpha$-CaCr$_{2}$O$_{4}$). The smaller separation between the triangular CrO$_{2}$  layers can enhance the interlayer interactions, which helps to stabilize the the long range ordering in the frustrated spin systems.

\begin{figure}
\includegraphics[width=0.5\textwidth]{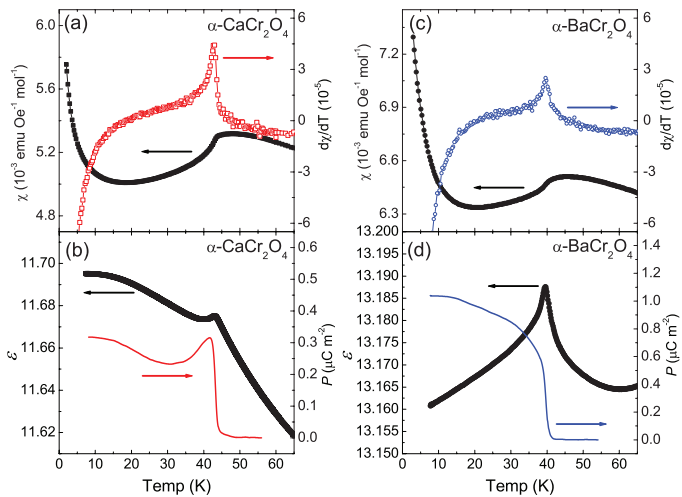}
\caption{\label{f7}
(Color online) Temperature dependence of the magnetic susceptibility ($\chi$), the corresponding ${\rm d}\chi/{\rm d}T$,  the dielectric constant ($\epsilon$) and the electric polarization ($P$) of $\alpha$-CaCr$_{2}$O$_{4}$  (a-b) and $\alpha$-BaCr$_{2}$O$_{4}$ (c-d).
}
\end{figure}

As shown in Fig. 7(b) and (d), like $\alpha$-SrCr$_{2}$O$_{4}$, the $\epsilon(T)$s of both $ \alpha$-CaCr$_{2}$O$_{4}$ and $\alpha$-BaCr$_{2}$O$_{4}$, show anomalous peaks with their corresponding maxima exactly located at their $T_N$ (43.0K and 39.5K respectively), coinciding well with the emerging of the long-range magnetic ordering. The corresponding difference between them is also considerable. The $\epsilon(T)$ of $\alpha$-CaCr$_{2}$O$_{4}$ exhibits a very weak peak at $T_N$, followed by a slow increase on further cooling, which behaves quite similar with $\alpha$-SrCr$_{2}$O$_{4}$.  As cooled from above $T_N$, the $P(T)$ arises sharply just below $T_N$, from zero to a maximum, then followed by a drop till 30K due to the emergent negative pyroelectric currents (not shown) as observed in $\alpha$-SrCr$_{2}$O$_{4}$ (shown in Fig. 4b). The behavior of $P(T)$ resembles that of $\alpha$-SrCr$_{2}$O$_{4}$ closely. The $P$ at 10K is 0.31 $\mu$C/m$^2$, slightly lower than $\alpha$-SrCr$_{2}$O$_{4}$.

On the other hand, around $T_N$(=39.5K), the peak in $\epsilon(T)$ of $\alpha$-BaCr$_{2}$O$_{4}$ is much more pronounced than the other two analogs. The $\epsilon(T)$ decreases monotonously with further cooling below $T_N$. The salient peak indicates a stronger ferroelectric transition. No negative pyroelectric current is observed on $\alpha$-BaCaCr$_{2}$O$_{4}$ samples below $T_N$. Correspondingly, $P(T)$  increases monotonously upon cooling as $T < T_N$. $P$(T=10K) reaches to 1 $\mu$C/m$^2$, about two times larger than the Sr or Ca analogs.

As for the magentoelectric effect in $\alpha$-CaCr$_{2}$O$_{4}$ and $\alpha$-BaCr$_{2}$O$_{4}$, both compounds exhibit a very weak magentoelectric coupling below $T_N$ as $\alpha$-SrCr$_{2}$O$_{4}$ (not shown). The MC values increase quasi-quadratically with respect to H, achieving the order of only $10^{-4}$ even at the highest H=9T, similar as in $\alpha$-SrCr$_{2}$O$_{4}$.

For a comprehensive understanding of the long-range magnetic ordering in the $\alpha$-MCr$_{2}$O$_{4}$ family, we plot all the specific heat data for all these three member compounds in Fig. 8. The clear $\lambda$-shaped anomaly peak is visible for each sample. There are no temperature shifts of the peak in $C_p(T)$ during the warming and cooling measuring process, which suggests the nature of second order in the magnetic transitions in this family.

\begin{figure}
\includegraphics[width=0.4\textwidth]{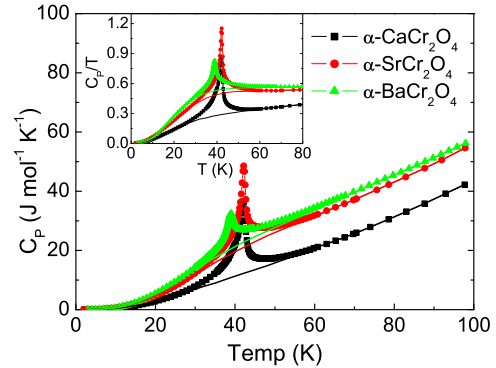}
\caption{\label{f8}
(Color online) Temperature dependence of the specific heat measured for three bulk samples of $\alpha$-phase CaCr$_{2}$O$_{4}$, SrCr$_{2}$O$_{4}$ and BaCr$_{2}$O$_{4}$. The dashed lines are the corresponding backgrounds to be subtracted in estimate the change of entropy by the specific heat anomalies, respectively. The corresponding $C_p(T)/T$ is plotted in the inset.
}
\end{figure}

To estimate the entropy removed by the long-range magnetic ordering which signaled by the peak of $C_p(T)$ anomaly around $T_N$, we use the same procedure as described earlier. The baseline for background subtraction (shown as dashed lines in Fig. 8) shifts vertically as M changes from the lightest element (Ca) to the heaviest Ba atom, consistent with the known Debye model for the specific heat of phonons. The different M atoms contribute the lattice specific heat, since Debye temperature $\theta_D \propto 1/\sqrt{W}$   where $W$ is the molecular weight.

The calculated change in entropy for $\alpha$-phase SrCr$_{2}$O$_{4}$,  CaCr$_{2}$O$_{4}$ and  BaCr$_{2}$O$_{4}$ is 3.6, 2.9,
2.0 \joulepermolekelvin, respectively, which is equivalent to 15.6\%, 12.6\%, 8.5\% of the total magnetic entropy respectively. Apparently, most of the missing entropy  is removed by low-dimensional short-range ordering well above $T_N$. The change in entropy decreases as the radius of interlayer nonmagnetic M ions increases. The increase in the spacing between CrO$_{2}$  layers can weaken the interlayer interaction, and drive the system closer to the ideal low-dimensionality and therefore the short-range ordering above $T_N$ becomes more dominant.

\section{Discussions}

The strong geometrically magnetic frustration and low-dimensionality dominate the physics in the $\alpha$-MCr$_{2}$O$_{4}$ family. The previous neutron diffraction experiments have revealed that there exists a quasi-$120\degree$ proper-screw-type helical spin structure with the magnetic propagation vector along $b$-axis emerging below $T_N$. This quasi-$120\degree$ spin picture is energetically favorite, consistent with the theory for 2D triangular lattice, which lead to the over-all similar low-temperature magnetic properties in this family.

The similar properties of magnetization, specific heat, dielectric constant, and ferroelectric polarization in these three members have been observed in our experiments. The very weak multiferroicity (compared with many known multiferroic materials \cite{LiuJM}) is confirmed below $T_N$, concomitant with the proper-screw-type spin ordering as $T<T_N$.  In order to better understand our observations, we analyzed our results in comparison with the existing theoretical models on the microscopic mechanism of multiferroicity. All the spins at Cr$^{3+}$ sites are equivalent (${\rm S}=3/2$). The macroscopic net exchange striction in helical magnetic structure averages to zero. So the magneto-striction cannot give net electric polarization. On the other hand, according to the inverse DM mechanism or the spin-current model, this proper-screw-type helical spins cannot induce any microscopic electric polarization.

A more plausible candidate, as proposed by Arima and Jia et al, \cite{Arima, Jia}  is  the $d$-$p$ hybridization modified by the spin-orbital interactions in partially filled $t_{2g}$ systems, which can induce the microscopic electric polarization along the bond direction. This kind of local polarization can be expressed as ${\bf P}_{ij} \propto ({\bm S}_i \cdot \hat{\bf {e}}_{ij}){\bm S}_i - ({\bm S}_j \cdot \hat{\bf {e}}_{ij}) {\bm S}_j $, where $\hat{\bf {e}}_{ij}$ is the unit vector connecting the neighboring sites $i$ and $j$. Although this term is oscillating and usually cancels out in zero in most crystals, in certain geometries like 2D-triangular systems with inversion-symmetry-breaking proper-screw-type helimagnetism, the net components along the modulation vector have been shown to be non-zero.\cite{Arima} This is the case in the $\alpha$-MCr$_{2}$O$_{4}$ family with proper-screw-type helimagnetism modulated along the $b$-axis. Therefore, the electric polarization is expected along $b$-axis, which is also consistent with the proposed $b$-axis polar order based on the symmetry analysis. In our measurements on crystal samples of $\alpha$-SrCr$_{2}$O$_{4}$, no ferroelectricity along $a$-axis can be observed, supporting the above theoretical analysis.

The effect of inter-layer interaction has not be considered in the Arima's theory. So far, we have observed that the multiferroicity in $\alpha$-MCr$_{2}$O$_{4}$ are enhanced as M changes from Ca to Ba with increasing separation between CrO$_{2}$  layers. The negative hump of pyroelectric currents are observed in $\alpha$-SrCr$_{2}$O$_{4}$ and $\alpha$-CaCr$_{2}$O$_{4}$ with small interlayer spacings, which may arise from the antiferroelectric correlations between the CrO$_2$ layers, and lead to the decrease of $P(T)$ on cooling below $T_N$. But this behavior no longer exists in $\alpha$-BaCr$_{2}$O$_{4}$, which has larger interlayer spacing than the other two members in this family. The interlayer exchange interaction in $\alpha$-MCr$_{2}$O$_{4}$ family accounts for this weak antiferroelectric interlayer coupling. The similar  interlayer antiferroelectric coupling has also been  discussed in some ACrO$_{2}$(A is alkali metal)  systems in detail by Seki et al.\cite{ACrO2}  Although the interlayer interaction stabilizes the long-range ordering and help to achieve higher $T_N$, it also favors the antiferroelectric coupling between triangular layers and greatly suppresses the total macroscopic polarization. The stronger multiferroicity seems to favor the more ideal low-dimensionality.

\section{Conclusion}

In this paper, the quasi-2D geometrically frustrated triangular magnetic system, $\alpha$-MCr$_{2}$O$_{4}$ (M=Ca, Sr, Ba), was systematically investigated, and the multiferroicity has been confirmed concomitant with the long range magnetic ordering below $T_N$ in this family. The possible microscopic mechanism of this spin-driven ferroelectricity is attributed to the Arima model for the proper screw type of helical magnetic ordering in a triangular antiferromagnetic lattice. The evolution of multiferroic properties with the increasing separation of the stacking CrO$_{2}$  layers is also discussed. Our present studies provide valuable information for the further search of new multiferroics in geometrically frustrated magnetic systems.

\begin{acknowledgments}

We wish to acknowledge Dr M.J Wang, W.L. Lee and C.C. Li for their technical support and helpful discussions. We also acknowledge the financial support from Academia Sinica and the National Science Council of Taiwan.

\end{acknowledgments}

\end{document}